\documentclass{svproc}
\usepackage{url}
\usepackage[ruled,linesnumbered,lined,commentsnumbered]{algorithm2e}

\SetAlFnt{\small}
\SetAlCapFnt{\small}
\SetAlCapNameFnt{\small}
\usepackage{algorithmic}
\algsetup{linenosize=\tiny}
\usepackage[tight]{subfigure}
\usepackage{graphicx}
\usepackage{multicol,lipsum}
\usepackage{amsfonts}
\usepackage{booktabs}
\usepackage{siunitx}
\usepackage{mathtools}
\usepackage{breqn}
\usepackage{placeins}
\usepackage{balance}

\usepackage{amsmath}

\SetKwRepeat{Do}{do}{while}

\begin{document}
\mainmatter              
\title{StreamNet: A DAG System with Streaming Graph Computing}
\titlerunning{StreamNet}  
%
\author{Zhaoming Yin\inst{1},\inst{3} \and Anbang Ruan\inst{2} \and Ming Wei\inst{2} 
\and Huafeng Li\inst{3} \and Kai Yuan\inst{3} \and Junqing Wang\inst{3} \and Yahui Wang\inst{3} 
\and Ming Ni\inst{3} \and Andrew Martin\inst{4} }
\authorrunning{Zhaoming Yin et al.} 
%
\tocauthor{Ivar Ekeland, Roger Temam, Jeffrey Dean, David Grove,
Craig Chambers, Kim B. Bruce, and Elisa Bertino}
\institute{StreamNet Chain LLC, Hangzhou, Zhejiang 310007, China,\\
\email{stplaydog@gmail.com},\\ WWW home page:
\texttt{http://www.streamnet-chain.com/}
\and
Octa Innovation, Beijing, 100036, China,\\
\email{ar,wm@8lab.cn}, \\ WWW home page:
\texttt{https://www.8lab.cn/}
\and
TRIAS Lab, Hangzhou, Zhejiang, 310008, China,\\
\email{lhf,yuankai,wangjunqing,wangyahui,ming.ni@trias.one}, \\ WWW home page:
\texttt{https://www.trias.one/}
\and
University of Oxford, Oxford, OX1 4BH, England,\\
\email{andrew.martin@cs.ox.ac.uk}, \\ WWW home page:
\texttt{https://www.cs.ox.ac.uk/people/andrew.martin/}
}

\maketitle              

\begin{abstract}
To achieve high throughput in the POW based blockchain systems, researchers proposed a series of methods, and DAG is one of the most active and promising fields.
We designed and implemented the StreamNet, aiming to engineer a scalable and endurable DAG system.
When attaching a new block in the DAG, only two tips are selected.
One is the `parent' tip whose definition is the same as in Conflux \cite{li2018scaling};
another is using Markov Chain Monte Carlo (MCMC) technique by which the definition is the same as IOTA \cite{popov2016tangle}.
We infer a pivotal chain along the path of each epoch in the graph, and a total order of the graph could be calculated without a centralized authority.
To scale up, we leveraged the graph streaming property; high transaction validation speed will be achieved even if the DAG is growing.
To scale out, we designed the `direct signal' gossip protocol to help disseminate block updates in the network, such that messages can be passed in the network more efficiently.
We implemented our system based on IOTA's reference code (IRI) and ran comprehensive experiments over the different sizes of clusters of multiple network topologies.
\
\end{abstract}

\section{Introduction}
Since bitcoin \cite{nakamoto2008bitcoin} has been proposed, blockchain technology has been studied for $10$ years.
Extensive adoptions of blockchain technologies was seen in real-world applications such as financial services with potential regulation challenges \cite{michael2018blockchain,tapscott2017blockchain},
supply chains \cite{korpela2017digital,tian2016agri,abeyratne2016blockchain}, 
health cares \cite{azaria2016medrec,yue2016healthcare} and IoT devices \cite{christidis2016blockchains}. 
The core of blockchain technology depends on the consensus algorithms applying to the open distributed computing world.
Where computers can join and leave the network, and these computers can cheat.

As the first protocol that can solve the so-called Byzantine general problem, 
the bitcoin system suffers from a low transaction rate with a transaction per second (TPS) of approximately $7$, 
and long confirmation time (about an hour). 
As more machines joined the network, they are competing for the privileges to attach the block (miners), 
which results in a massive waste of electric power. 
While skyrocketing fees are paid to make sure the transfers of money will be placed in the chain. 
On par, there are multiple proposals to solve the low transaction speed issue. 
One method intends to solve the speed problem without changing the chain data structure, 
for instance, segregated witness \cite{lombrozo2015segregated} or off-chain technologies such as lightning network \cite{poon2016bitcoin} or plasma \cite{poon2017plasma}. 
Another hard fork way changed the bitcoin protocol, such as the bitcoin cash tries to improve the throughput of the system by enlarging the data size of each block from $1$ Mb to $4$ Mb.

To minimize the computational cost of POW, 
multiple organizations have proposed a series of proof of stake method (POS) 
\cite{duffield2018dash,tron2018,david2017ouroboros,wood2014ethereum,goodman2014tezos} 
to make sure that those who have the privilege to attach the block proportional to their token shares. 
Another idea targeting at utilizing the power in POW to do useful and meaningful tasks such as training machine learning models are also proposed \cite{matthew2017aion}. 
Besides, inspired by the PBFT algorithm \cite{castro1999practical} and a set of related variations, 
the so-called hybrid (or consortium) chain was proposed. 
The general idea is to use a two-step algorithm; the first step is to elect a committee; 
the second step is collecting committee power to employ PBFT for consensus. Bitcoin-NG \cite{eyal2016bitcoin} is the early adopter of this idea, 
which splits the blocks of bitcoin into two groups: 
for master election and another for regular transaction blocks. 
Honey-badger \cite{miller2016honey} is the system that first introduced the consensus committee; 
it uses predefined members to perform the PBFT algorithm to reach consensus. 
The Byzcoin system \cite{kogias2016enhancing} brought forth the idea of POW for the committee election and uses a variation of PBFT called collective signing for speed purposes. 
The Algorand \cite{gilad2017algorand} utilizes a random function to elect a committee and use this committee to commit blocks anonymously, 
and the member of the committee only has one chance to commit block. 
Other popular systems include Ripple \cite{schwartz2014ripple}, Stellar \cite{mazieres2015stellar} and COSMOS \cite{kwon2016cosmos} etc. 
All these systems have one common feature, the split of layers of players in the network, which results in the implementation complexity.
While the methods above are aiming to avoid side chains, another thread of effort is put on using a direct acyclic graph(DAG) to merge side chains. 
The first-ever idea comes with growing the blockchain with trees instead of chains \cite{sompolinsky2013accelerating}, 
which results in the well-known GHOST protocol \cite{sompolinsky2015secure}. 
If one block links to $\geq 2$ previous blocks, then the data structure grows like a DAG instead of tree \cite{lewenberg2015inclusive}, 
SPECTRE \cite{sompolinsky2016spectre} and PHANTOM \cite{sompolinskyphantom} are such type of systems.
Byteball \cite{churyumov2016byteball} is the system that constructs the main chain, and leverage this main chain to help infer the total order, nonetheless, 
the selection of the main chain is dependent on a role called to witness, which is purely centralized. 
Conflux is an improvement of the GHOST based DAG algorithm, 
which also utilizes the pivotal (main) chain without the introduction of witness and claim to achieve $6000$ of TPS in reality \cite{li2018scaling}. 
IOTA tried to avoid the finality of constructing a linear total order by introducing the probabilistic confirmation in the network \cite{popov2016tangle}. 
As mentioned earlier, the systems are permissionless chains; 
in the permission chains, DAG technology is also applied. HashGraph \cite{baird2016swirlds} is the system that utilizes the $gossip\ on\ gossip$ algorithm to propagate the block graph structure, 
and achieve the consensus by link analysis in the DAG, this method is proved to be Byzantine fault-tolerant and does not rely on voting. 
Blockmainia \cite{danezis2018blockmania} is based on the original PBFT design, but its underlying log is DAG-based. 
Some of the side chain methods also borrow the idea of DAG, such as nano \cite{lemahieu2018nano} and VITE \cite{liuvite}. 
These systems, in reality, rely on centralized methods to maintain their stability.

Social network analysis has widely adopted the method of streaming graph computing \cite{ediger2011tracking,green2012fast,ediger2012stinger}, 
which deals with how to quickly maintain information on a temporally or spatially changing graph without traversing the whole graph. 
We view the DAG-based method as a streaming graph problem, which is about computing the total order and achieving consensus without consuming more computing power. 
In distributed database systems, the problem of replicating data across machines is a well-studied topic \cite{demers1988epidemic}. 
Due to the bitcoin network's low efficiency, there are multiple ways to accelerate the message passing efficiency \cite{klarmanbloxroute}. 
However, they did not deal with network complexity. 
We viewed scaling the DAG system in the network of growing size and topological complexity as another challenging issue and proposed our gossip solution. 
This paper's main contribution is how to utilize the streaming graph analysis methods and new gossip protocol to enable real decentralized, and stabilized growing DAG system.


\section{Basic design}

\subsection{Data structure}

\begin{figure}[!ht]
\begin{center}
\includegraphics[width=0.65\textwidth]{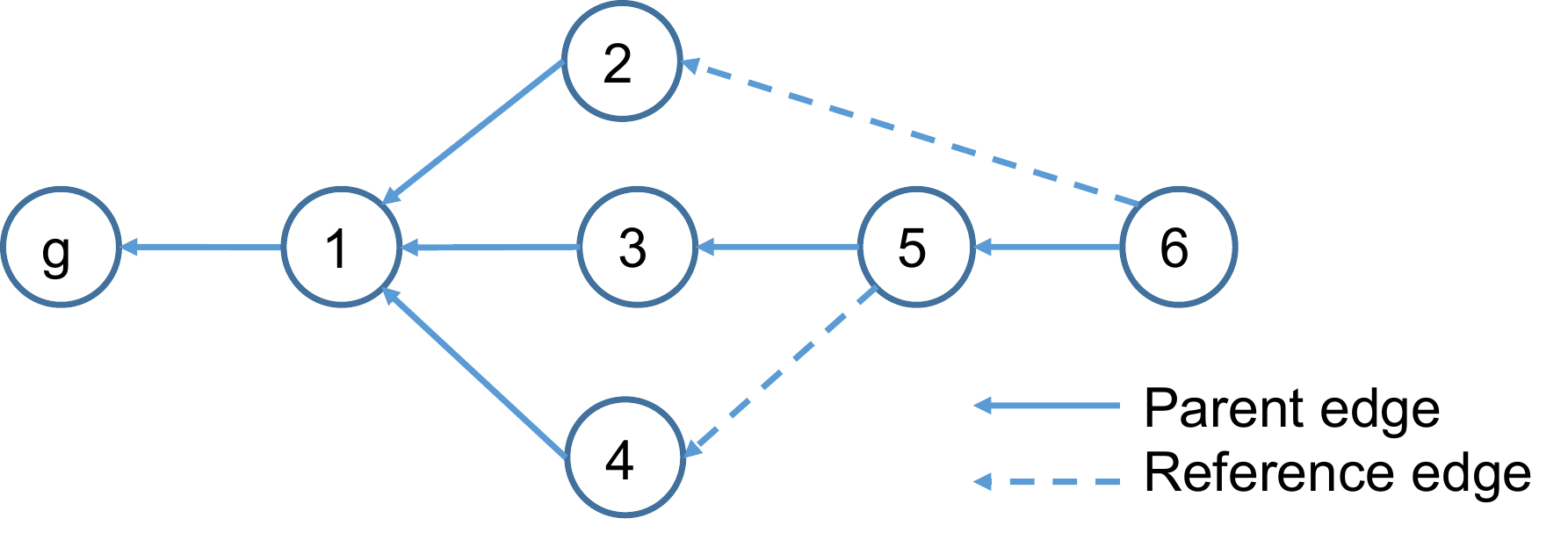}
  \caption{
    Example of the StreamNet data structure.
   }
\label{simple_sn}
\end{center}
\end{figure}

The local state of a node in the StreamNet protocol is a direct acyclic graph (DAG) $G = <B,g,P,E>$. $B$ is the set of blocks in $G$.
$g \in G$ is the genesis block. 
For instance, vertex $g$ in Figure~\ref{simple_sn} represents the Genesis block. 
$P$ is a function that maps a block $b$ to its parent block $P(b)$. 
Specially, $P(g) = \perp$. In Figure~\ref{simple_sn}, parent relationships are denoted by solid edges. 
Note that there is always a parent edge from a block to its parent block (i.e., $\forall b \in B$, $b, P(b)> \in E$). 
$E$ is the set of directly reference edges and parent edges in this graph. $e = <b,b'> \in E$ is an edge from the block $b$ to the block $b'$, 
which means that $b'$ happens before $b$. 
For example in Figure~\ref{simple_sn}, vertex $1$ represents the first block, 
which is the parent for the subsequent block $2$, $3$ and $4$. 
Vertex $5$ has two edges; one is the parent edge pointing to $3$, another is reference edge pointing to $4$. 
When a new block is not referenced, it is called a tip. 
For example, in Figure~\ref{simple_sn}, block $6$ is a tip. 
All blocks in the StreamNet protocol share a predefined deterministic hash function Hash that maps each block in $B$ to a unique integer id. 
It satisfies that $\forall {b} \neq {b'}$, Hash($b$) $\neq$ Hash($b'$).

\subsection{StreamNet Architecture}

\IncMargin{1em}
\begin{algorithm}
\SetKwData{Left}{left}\SetKwData{This}{this}\SetKwData{Up}{up}
\SetKwFunction{Union}{Union}\SetKwFunction{FindCompress}{FindCompress}
\SetKwInOut{Input}{input}\SetKwInOut{Output}{output}

\KwIn{ Graph $G=<B, g, P, E>$ }

    \While {Node is running}{
        \uIf{Received $G' = <B', g, P', E'> $}{
            $G'' \gets <B \cup B', g, P \cup P', E \cup E' > $\;
            \uIf{$G \neq G''$ } {
                $G \gets G''$ \;
                Broadcase updated G to neighbors \;
            }
        } 

        \uIf{Generate block $b$}{
            $a \gets Pivot(G, g) $ \;
            $r \gets MCMC(G, g) $ \;
            $G \gets <B \cup b, g, P \cup <b, a>, E \cup <b, a> \cup <b, r> >$ \;
            Broadcase updated G to neighbors \;
        } 
    }

\caption{{ StreamNet node main loop.}}
\label{algo:main_loop}
\end{algorithm}
\DecMargin{1em}

\begin{figure}[!ht]
\begin{center}
\includegraphics[width=0.85\textwidth]{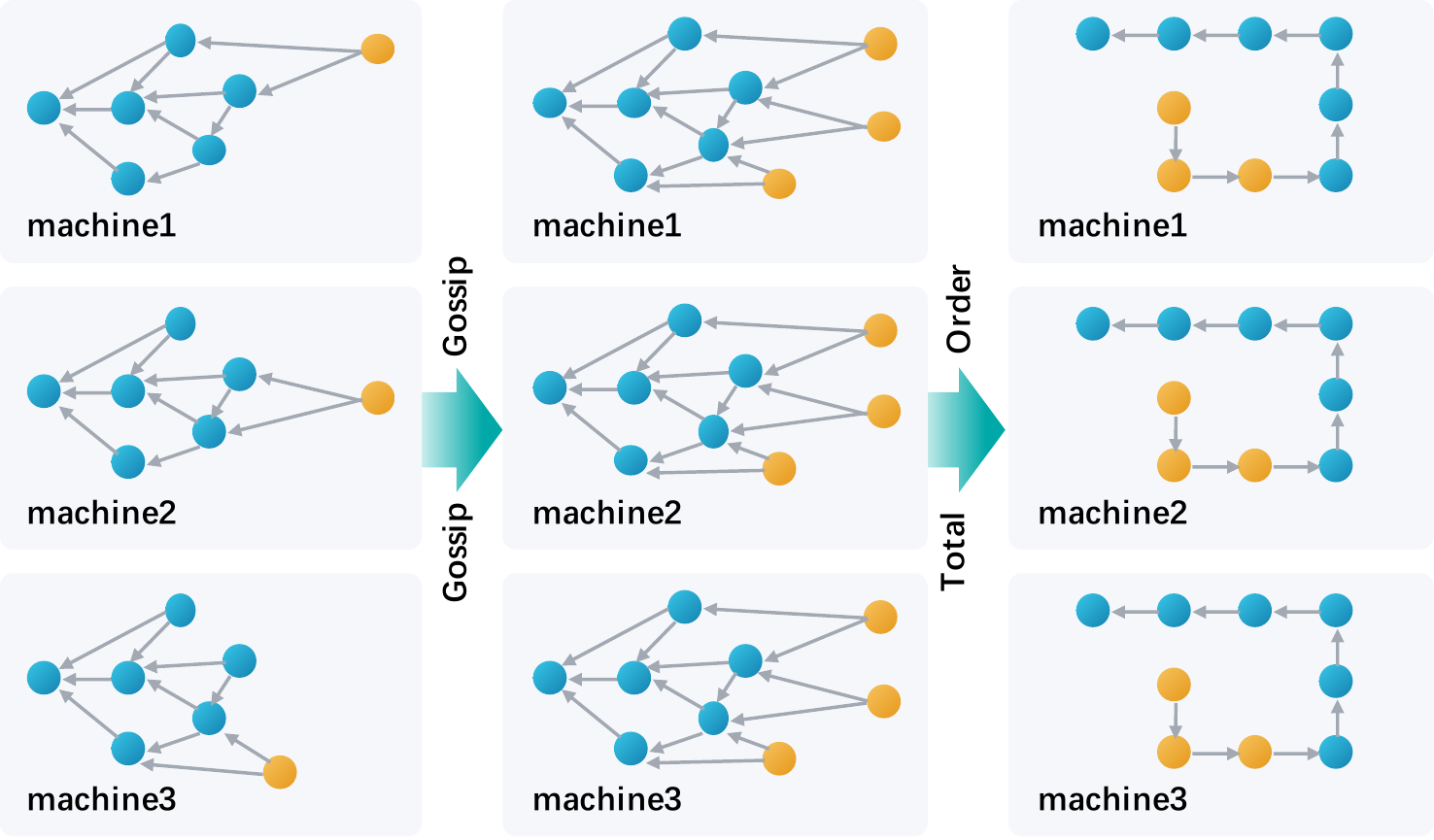}
  \caption{
    StreamNet architecture.
   }
\label{architecture}
\end{center}
\end{figure}

\begin{figure}[!ht]
\begin{center}
\includegraphics[width=0.95\textwidth]{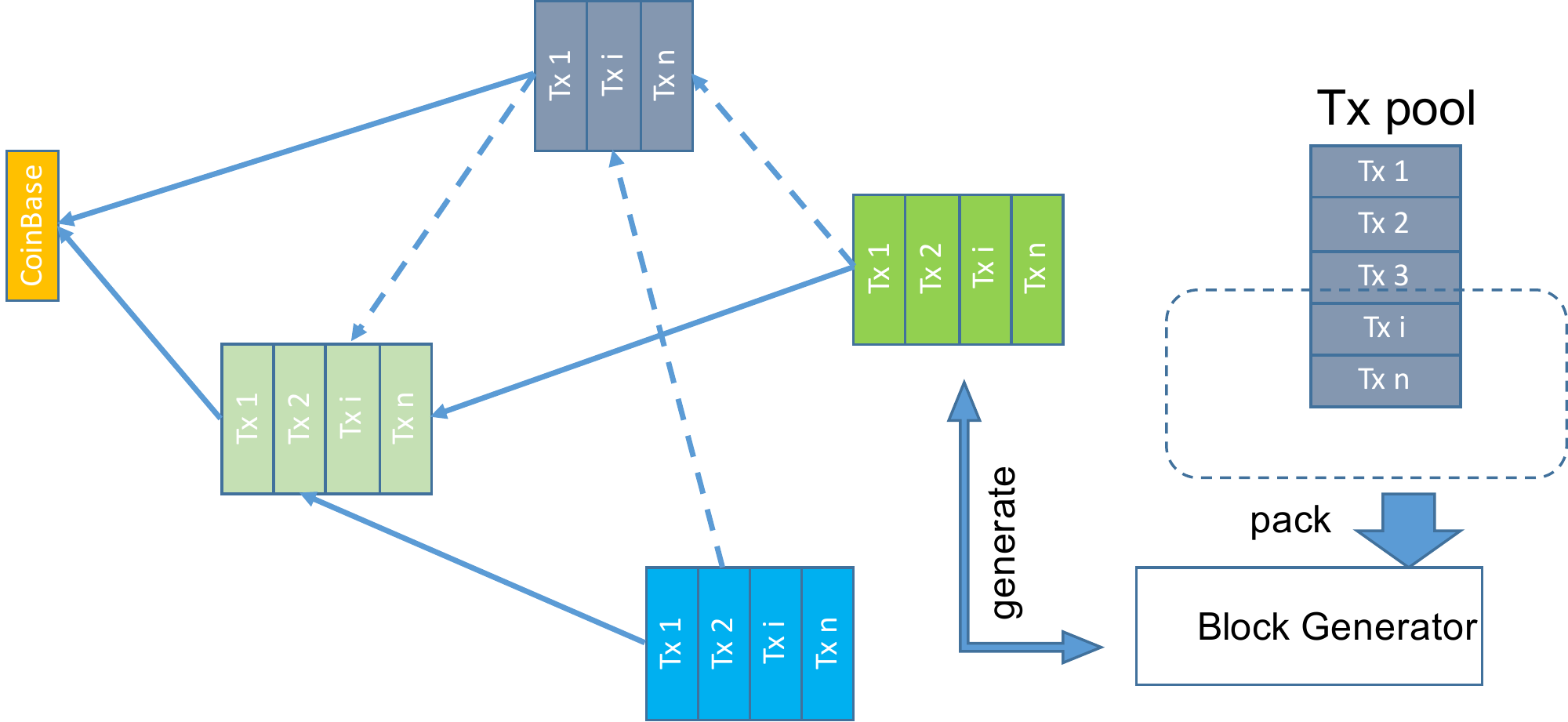}
  \caption{
    One node in StreamNet protocol.
   }
\label{node}
\end{center}
\end{figure}

Figure~\ref{architecture} presents the architecture of StreamNet; 
it is consists of multiple StreamNet machines. 
Each StreamNet machine will grow its DAG locally and will broadcast the changes using the gossip protocol. 
Eventually, every machine will have a unified view of DAG.
By calling the total ordering algorithm, every machine can sort the DAG into a total order, 
and the data in each block can have a relative order regardless of their local upload time. 
Figure~\ref{node} shows the local architecture of StreamNet. In each StreamNet node, 
there will be a transaction pool accepting the transactions from the HTTP API. 
Moreover, there will be a block generator to pack a certain amount of transactions into a block, 
and it firstly finds a parent and reference block to attach the new block to, based on the hash information of these two blocks and the metadata of the block itself, 
it will then perform the proof of work (POW) to calculate the nonce for the new block. 
Algorithm~\ref{algo:main_loop} summarize the server logic for a StreamNet node. 
In the algorithm, the way to find parent block is by $Pivot(G, g)$. 
Furthermore, the way to find a reference block is by calling $MCMC(G, g)$, 
which is the Markov Chain Monte Carlo (MCMC) random walk algorithm \cite{popov2016tangle}. 
The two algorithms will be described in the later section.

\subsection{Consensus protocol}
Based on the predefined data structure, to present the StreamNet consensus algorithm, 
we firstly define several utility functions and notations, which is a variation from the definition in the Conflux paper \cite{li2018scaling}. 
Chain() returns the chain from the genesis block to a given block following only parent edges. 
$\overline{Chain(G,b)}$ returns all blocks except those in the chain. 
Child() returns the set of child blocks of a given block. Sibling() returns the set of siblings of a given block.
Subtree() returns the subtree of a given block in the parental tree.
Before() returns the set of blocks that are immediately generated before a given block. 
Past() returns the set of blocks generated before a given block (but including the block itself). 
After() returns the set of blocks that are immediately generated after a given block. 
Later() returns the set of blocks generated after a given block (but including the block itself). 
SubGraph() returns the subgraph by removing blocks and edges except for the initial set of blocks. 
ParentScore() presents the weight of blocks, and each block has a score when referenced as a parent. 
Score() presents the weight of blocks, and each block achieves a score when attaching to the graph. 
TotalOrder() returns the `flatten' order inferred from the consensus algorithm. 
Figure~\ref{allMethods} represents the definition of these utility functions.

\begin{figure}
\begin{flalign*}
 &\fbox{G = $<B,g,P,E>$} \\
 &Chain(G,b) =
 \begin{cases}
  g         & \text{b = g} \\
  Chain(G,P(b))   & \text{otherwise}
 \end{cases} \\
 & \overline{Chain(G,b)} = \{ b' | b' \in B, b' \notin Chain(G,b) \} \\
  &Child(G,b) = \{ b'| P(b') = b \} \\
  &Sibling(G,b) = Child(G,P(b)) \\
  &SubTree(G,b) = (U_{i\in Child(G,b)}Substree(G,i)) \cup \{b\} \\
  &Before(G,b) = \{b'|b' \in B, <b,b'> \in E \} \\
  &Past(G,b) = (U_{i\in Before(G,b)}Past(G,i)) \cup \{b\} \\
  &After(G,b) = \{b'|b' \in B, <b',b> \in E \} \\
  &Later(G,b) = (U_{i\in After(G,b)}Later(G,i)) \cup \{b\} \\
  &SubGraph(G,B') = <B', P', E'> | \\
  & \forall <b, b'> \in E', b \subset B' \& b' \subset B'\\
  &ParentScore(G,b) = |SubTree(G,b)| \\
  &Score(G,b) = |Later(G,b)| \\
  &TotalOrder(G) = StreamNetOrder(G,Pivot(G,g)) 
\end{flalign*}

  \caption{The Definitions of Chain(), Child(), Sibling(), Subtree(), Before(), Past(), After(), Later(), SubGraph(), ParentScore(), Score(), and TotalOrder(). }
\label{allMethods}
\end{figure}

\subsubsection{Parent tip Selection by pivotal chain}

\IncMargin{1em}
\begin{algorithm}
\SetKwData{Left}{left}\SetKwData{This}{this}\SetKwData{Up}{up}
\SetKwFunction{Union}{Union}\SetKwFunction{FindCompress}{FindCompress}
\SetKwInOut{Input}{input}\SetKwInOut{Output}{output}

\KwIn{ The local state $G$ = $<B,g,P,E>$  and a starting block $b \in B$ }
\KwOut{ A random tip $t$ }

$t \leftarrow b$

\Do { Score(G,t) != 0} {
    \For {$b' \in Child(G,t)$} {
        $P_{bb'} = \frac{e^{\alpha Score(G,b')}}{\Sigma_{z:z \rightarrow b}e^{\alpha Score(G,z)}}$
    }
    $t \leftarrow $ choose $b''$ by $P_{bb''}$
}

\Return{$t$} \;

\caption{{\sc MCMC($G$, $b$).}}
\label{algo:mcmc}
\end{algorithm}
\DecMargin{1em}

\begin{figure}[!ht]
\begin{center}
\includegraphics[width=0.75\textwidth]{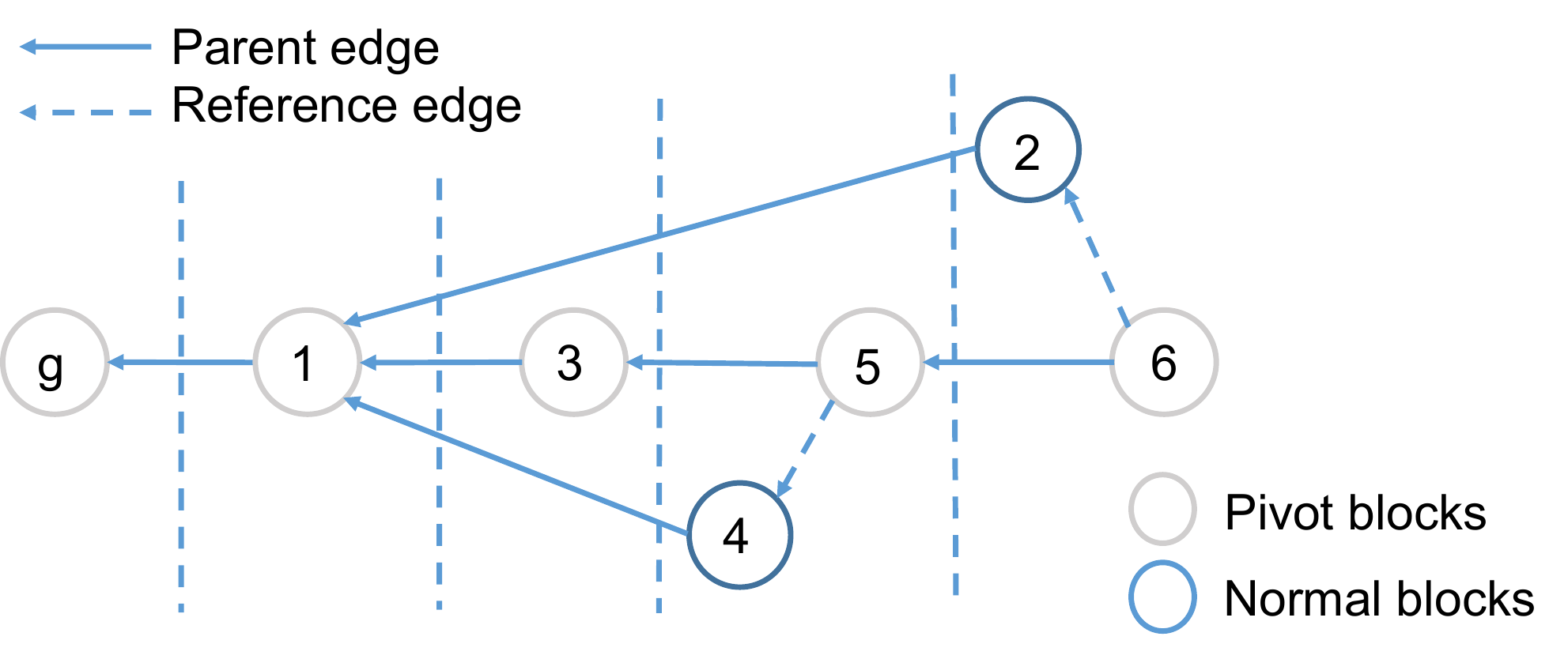}
  \caption{
    An example of total order calculation.
   }
\label{total_order}
\end{center}
\end{figure}

The algorithm Algorithm~\ref{algo:getPivot} presents our pivot chain selection algorithm(i.e., the definition of $Pivot(G, b)$). 
Given a StreamNet state $G$, Pivot($G$,$g$) returns the last block in the pivoting chain starting from the genesis block $g$. 
The algorithm recursively advances to the child block, whose corresponding sub-tree has the most significant number of children. 
Which is calculated by $ParentScore(G, b)$ When there are multiple child blocks with the same score, 
the algorithm selects the child block with the largest block hash. 
The algorithm terminates until it reaches a tip. Each block in the pivoting chain defines an epoch, 
the nodes in DAG that satisfy Past($G$,$b$) - Past($G$,$p$) will belong to the epoch of block $b$. 
For example, in Figure~\ref{total_order}, the pivoting chain is $<g, 1, 3, 5, 6>$, and the epoch of block $5$ contains two blocks $4$ and $5$.

\IncMargin{1em}
\begin{algorithm}
\SetKwData{Left}{left}\SetKwData{This}{this}\SetKwData{Up}{up}
\SetKwFunction{Union}{Union}\SetKwFunction{FindCompress}{FindCompress}
\SetKwInOut{Input}{input}\SetKwInOut{Output}{output}

\KwIn{ The local state $G$ = $<B,g,P,E>$  and a starting block $b \in B$ }
\KwOut{ The tip in the pivot chain }

\Do { Child(G,b) != 0} {
  $b'$ $\gets$ Child($G,b$)  \;
  $tmpMaxScore$ $\gets$ -1 \;
  $tmpBlock$ $\gets$ $\perp$ \;
  \For {$b' \in Child(G,b)$} {
    $pScore$ $\gets$ ParentScore($G, b'$) \;
    \If { $score$ $>$ $tmpMaxScore$ $||$ ($score$ = $tmpMaxScore$ \text{and Hash($b'$ ) $<$ Hash($tmpBlock$)}} {
      $tmpMaxScore$ $\gets$ $pScore$ \;
      $tmpBlock$ $\gets$ $b'$ \; 
    }
  }
  $b$ $\gets$ $tmpBlock$ \;
}

\Return{$b$} \;

\caption{{\sc pivot($G$, $b$).}}
\label{algo:getPivot}
\end{algorithm}
\DecMargin{1em}

\subsubsection{Reference tip selection by MCMC}

The tip selection method by using Monte Carlo Random Walk (MCMC) is as Algorithm~\ref{algo:mcmc} shows. 
Each random walk step, starting from the genesis, 
will choose a child to jump to, and the probability of jumping from one block to the next block will be calculated using the formula in the algorithm. 
$\alpha$ in the formula is a constant that is used to scale the randomness of the MCMC function, 
the smaller it is, the more randomness will be in the MCMC function. The algorithm returns until it finds a tip.

\subsubsection{Total Order}
Algorithm~\ref{algo:conflux_order} defines StreamNetOrder(), which corresponds to our block ordering algorithm.
Given the local state $G$ and a block $b$ in the pivoting chain,
StreamNetOrder($G$, $b$) returns the ordered list of all blocks that appear in or before the epoch of $b$. 
Using StreamNetOrder(), the total order of a local state $G$ is defined as TotalOrder($G$).
The algorithm recursively orders all blocks in previous epochs(i.e., the epoch of $P(b)$ and before). 
It then computes all blocks in the epoch of $b$ as $B_\Delta$. It topologically sorts all blocks in $B_\Delta$ and appends it into the result list. 
The algorithm utilizes a unique hash to break ties. In Figure~\ref{total_order}, the final total order is $<g, 1, 3, 4, 5, 2, 6>$.

\IncMargin{1em}
\begin{algorithm}
\SetKwData{Left}{left}\SetKwData{This}{this}\SetKwData{Up}{up}
\SetKwFunction{Union}{Union}\SetKwFunction{FindCompress}{FindCompress}
\SetKwInOut{Input}{input}\SetKwInOut{Output}{output}

\KwIn{ The local state $G$ = $<B,g,P,E>$  and a tip block $b \in B$ }
\KwOut{ The block list of total top order starting from Genesis block to the giving block $b$ in $G$ }

$L = \perp$

\Do { $b$ != $g$} {
  $p$ $\gets$ Parent($G,b$)  \;
  $B_\Delta$ $\gets$ Past($G$,$b$) - Past($G$,$p$) \;
  \Do { $B_\Delta$ $\neq$ 0 } {
      $G'$ $\gets SubGraph(B_\Delta) $ \;
    $B'_\Delta$ $\gets$ \{x $||$ Before($G'$,$x$) = 0\} \;
    \text{Sort all blocks in $B'_\Delta$ in order as $b'_1,b'_2,...,b'_k$} \\
      \text{such that $\forall$1$\leq i \leq j \leq k$, Hash($b'_i$) $\leq$ Hash($b'_j$)} \;  
    $L$ $\gets$ $L + b'_1 + b'_2 + ... + b'_k$ \;
    $B_\Delta$ $\gets$ $B_\Delta$ - $B'_\Delta$ \;
  }
  $b$ = $p$ \;
}

\Return{$L$} \;

\caption{{\sc StreamNetOrder($G$, $b$).}}
\label{algo:conflux_order}
\end{algorithm}
\DecMargin{1em}

\subsection{The UTXO model}

\begin{figure}[!ht]
\begin{center}
\includegraphics[width=0.85\textwidth]{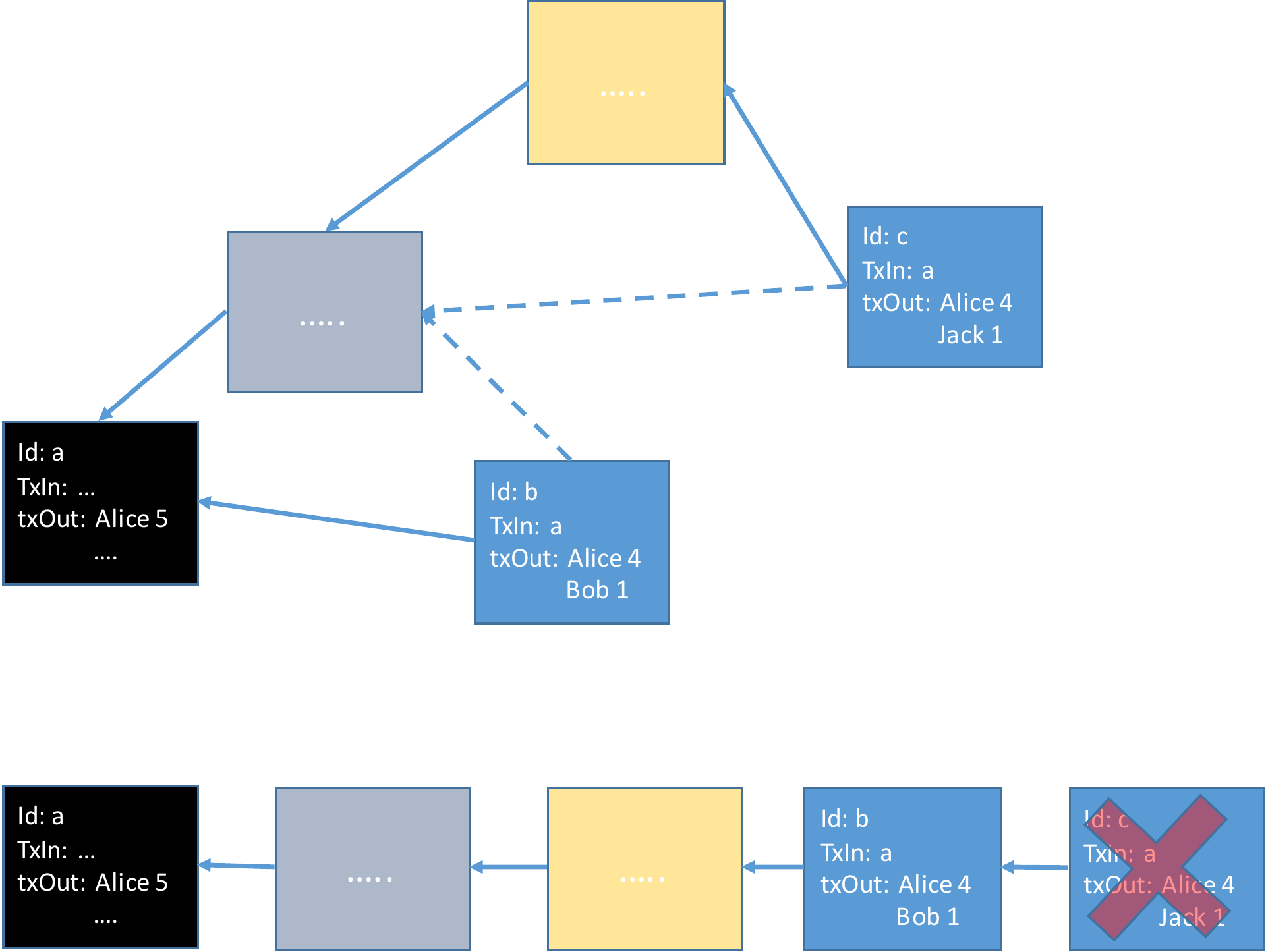}
  \caption{
    An example of UTXO.
   }
\label{utxo}
\end{center}
\end{figure}

In StreamNet, the transactions utilize the unspent transaction out (UTXO) model, 
which is the same as in Bitcoin. In the confirmation process, 
the user will call $TotalOrder$ to get the relative order of different blocks, 
and the conflict content of the block will be eliminated if the order of the block is later than the one conflicting with it in the total order. 
Figure ~\ref{utxo} shows the example of the storage of UTXO in StreamNet and how the conflict is resolved. 
Two blocks referenced the same block with Alice having five tokens and constructing the new transaction out, 
representing the transfer of token to Bob and Jack, respectively. 
However, after calling $totalOrder()$, the Bob transfer block precedes the Jack transfer block; thus, the next block will be discarded.

\subsection{Gossip Network}
In the bitcoin and IOTA network, the block information is disseminated in a direct mail way \cite{demers1988epidemic}. 
Suppose there are $N$ nodes and $L$ links in the network, 
for a block of size $B$, to spread the information of it, the direct mail algorithm will have a total complexity of $O(LB)$. 
Moreover, the average complexity for a node will be $O(\frac{LB}{N})$ In the chain based system, 
and this is fine because the design of the system already assumes that the transaction rate will below. 
However, in the DAG-based system, this type of gossip manner will result in low scalability due to the high throughput of the block generation rate and will result in network flooding. 
What is worse, consider the heterogeneously and long diameters of network topology, the convergence of DAG will take a long time, which will cause the delay of confirmation time of blocks.

\subsection{Differences with other DAG protocols}
Here, we mainly compare the difference of our protocol with two mainstream DAG-based protocols. One is IOTA, and another is Conflux.

\subsubsection{IOTA}
The major difference with IOTA is in three points:
\begin{itemize}
  \item Firstly, the IOTA tip selection algorithm's two tips are all randomly chosen, and ours is one deterministic which is for the total ordering purposes and one by random which is for maintaining the DAG property;
  \item Secondly, the IOTA consensus algorithm is not purely decentralized, it relies on a central coordinator to issue milestones for multiple purposes, and our algorithm does not depend on such a facility.
  \item Lastly, in IOTA, there is no concept of total order,  and there are three ways to judge if a transaction is confirmed:
  \begin{itemize}
    \item The first way is that the common nodes covered by all the tips are considered to be fully confirmed;
    \item All transactions referenced by the milestone tip are confirmed.
    \item The third way is to use MCMC. Call $N$ times to select a tip using the tip selection algorithm.
      If this tip references a block, its credibility is increased by 1.
      After $N$ selections have been cited $M$ times, then the credibility is $M / N$.
  \end{itemize}
\end{itemize}

\subsubsection{Conflux}
The major difference with Conflux is in two points:
\begin{itemize}
  \item Firstly, Conflux will approve all tips in the DAG along with the parent, which is much more complicated than our MCMC based two tip method. Moreover, when the width of DAG is high, there will be much more space needed to maintain such data structure.
  \item Secondly, the Conflux total ordering algorithm advances from genesis block to the end while StreamNet advances in the reverse direction. This method is one of the major contributions to our streaming graph-based optimizations, which will be discussed in the next chapter. In Conflux paper, there is no description of how to deal with the complexity paired with the growing graph.
\end{itemize}

\subsection{Correctness}

\subsubsection{Safety \& Liveness}
Because StreamNet utilizes the GHOST rule to select the pivoting chain, 
which is the same as in Conflux. 
Thus, it shares the same safety and correctness property as Conflux. 
Although the choice of reference chain in StreamNet is different from Conflux, it only affects the inclusion rate, which is the probability of a block to be included in the total order.

\subsubsection{Confirmation}

According to Theorem 10 in \cite{sompolinsky2015secure} and the deduction in \cite{li2018scaling}, given a period of $[t-d, t]$, and block $b$ in pivot chain in this period, the chance of $b$ kicked out by its sibling $b'$ is no more than $Pr(b_{drop})$ in (1). Which is the same as in Conflux.

\begin{equation}
     Pr(b_{drop}) \leq \sum_{k=0}^{n-m}{\zeta_kq^{n-m-k+1}} + \sum_{k=n-m+1}^{\infty}{\zeta_k} \\
     \zeta_k = e^{-q\lambda_{h}t \frac{(-q\lambda_{h}t)^k}{k!}}
\end{equation}

Followed by the definitions in Conflux paper \cite{li2018scaling},
in (1), $n$ is the number of blocks in the subtree before $t$, 
$m$ is the number of blocks in subtree of $b'$ before $t$. 
$\lambda_{h}$ is an honest node's block generation rate. 
$q (0 \leq q \leq 1)$ is the attacker's block generation ratio with respect to $\lambda_{h}$. 
From the equation, we can conclude that with the time $t$ goes, 
the chance of a block $b$ in the pivoting chain to be reverted is decreased exponentially.


\section{Optimization Methods}
One of the biggest challenges to maintain the stability of the DAG system is that, as the local data structure grows, the graph algorithms ($Pivot()$, $MCMC()$, $StreamNetOrder()$), relies on some of the graph operators that need to be recalculated for every newly generated block, which is very expensive. Table~\ref{tab:properties} list all the expensive graph operators that are called. Suppose the depth of the pivoting chain is $d$, then we give the analysis of complexity in the following way. $ParentScore()$ and $Score()$ rely on the breadth-first search ($BFS$), and the average $BFS$ complexity would be $O(|B|)$, and for each MCMC() and Pivot() called the complexity would be in total $O(|B|^2)$ in both of these two cases. The calculation of $Past()$ also relies on the $BFS$ operator, in the StreamNetOrder() algorithm, the complexity would be accrued to $O(|B| * d)$. TopOrder() is used in sub-order ranking the blocks in the same epoch. It is the classical topological sorting problem, and the complexity in the StreamNetOrder() would be $O(|B|)$.

\begin{table}[]
\caption {Analysis of Graph properties calculation} \label{tab:properties}
\begin{center}
\begin{tabular}{|l|l|l|l|}
\hline
Graph Property  & Algorithm used & Complexity & Tot \\ \hline
ParentScore(G, b)  & Pivot()   & $O(|B|)$    & $O(|B|^2)$ \\ \hline
Score(G, b)   & MCMC()   & $O(|B|)$    & $O(|B|^2)$ \\ \hline
Past(G,b) - Past(G,p) & StreamNetOrder() & $O(|B|)$    & $O(|B|*d)$ \\ \hline
TopOrder(G, b)  & StreamNetOrder() & $O(|B|)$    & $O(|B|)$ \\ \hline
\end{tabular}
\end{center}
\end{table}

Considering new blocks are generated and merged into the local data structure in a streaming way. The expensive graph properties could be maintained dynamically as the DAG grows. Such that the complexity of calculating these properties would be amortized to each time a new block is generated or merged. In the following sections, we will discuss how to design streaming algorithms to achieve this goal.

\subsection{Optimization of Score() and ParentScore() }
\IncMargin{1em}
\begin{algorithm}
\SetKwData{Left}{left}\SetKwData{This}{this}\SetKwData{Up}{up}
\SetKwFunction{Union}{Union}\SetKwFunction{FindCompress}{FindCompress}
\SetKwInOut{Input}{input}\SetKwInOut{Output}{output}

\KwIn{ Graph $G$, Block $b$, Score map $S$}
\KwOut{ Updated score map $S$ }

    $Q = [b]$ \;
    $visited = \{\}$ \;
    \While{ $Q != \O$} {
        $b'$ = $Q.pop()$ \;
        \For { $b'' \in Before(G, b')$ } {
            \uIf{$b'' \notin visited \land b'' != \perp$} {
                $Q.append(b'')$ \;
                $visited.add(b'')$ \;
            }
        }
        $S[b'] ++$ \;
    }

\Return{$S$} \;

\caption{{\sc UpdateScore($G$, $b$).}}
\label{algo:update_score}
\end{algorithm}
\DecMargin{1em}

In the optimized version, the DAG will have a map that keeps the score of each block. Once there is a new generated/merged block, it will trigger the BFS based UpdateScore() algorithm to update the block's scores in the map that are referenced by the new block. The skeleton of the UpdateScore() algorithm is as Algorithm~\ref{algo:update_score} shows.

\subsection{Optimization of Past(G,b) - Past(G,p)}

\IncMargin{1em}
\begin{algorithm}
\SetKwData{Left}{left}\SetKwData{This}{this}\SetKwData{Up}{up}
\SetKwFunction{Union}{Union}\SetKwFunction{FindCompress}{FindCompress}
\SetKwInOut{Input}{input}\SetKwInOut{Output}{output}

\KwIn{ Graph $G$, Block $b$, covered block set $C$}
\KwOut{ diff set $D \gets Past(G,b) - Past(G,p)$ }

    $D = {\O}$ \;
    $Q \gets [b]$ \;
    $visited = \{b\}$ \;
    $p$ = $Parent(G, b)$ \;
    \While{ $Q != \O$} {
        $b'$ = $Q.pop()$ \;
        \For { $b'' \in Before(G, b')$ } {
            \uIf{$IsCovered(G, p ,b'',C) \land b'' != \perp$} {
                $Q.append(b'')$ \;
                $visited.add(b'')$ \;
            }
        }
        $D.add(b')$ \;
        $C.add(b')$ \;
    }

\Return{$D$} \;

\caption{{\sc GetDiffSet($G$, $b$, $C$).}}
\label{algo:get_diff_set}
\end{algorithm}
\DecMargin{1em}

\IncMargin{1em}
\begin{algorithm}
\SetKwData{Left}{left}\SetKwData{This}{this}\SetKwData{Up}{up}
\SetKwFunction{Union}{Union}\SetKwFunction{FindCompress}{FindCompress}
\SetKwInOut{Input}{input}\SetKwInOut{Output}{output}

\KwIn{ Graph $G$, Block $b'$, parent $p$, covered block set $C$}
\KwOut{true if covered by parent, else false }
    $Q \gets [b']$ \;
    $visited = \{b\}$ \;
    \While{ $Q != \O$} {
        $b''$ = $Q.pop()$ \;
        \For { $ t \in Child(G, b'')$ } {
            \uIf{$t=p$}{
                return true \;
            }
            \uElseIf{$t \notin visited \land t \notin C$} {
                $Q.add(t)$ \;
                $visited.add(t)$ \;
            }
        }
    }

\Return{false} \;

\caption{{\sc IsCovered($G$, $p$, $b'$, $C$).}}
\label{algo:is_covered}
\end{algorithm}
\DecMargin{1em}

We abbreviate the Past(G,b) - Past(G,p) to calculate $B\delta$ as GetDiffSet(G,b,C) which is shown in the Algorithm~\ref{algo:get_diff_set}. This algorithm is, in essence, a dual-direction $BFS$ algorithm. Starting from the block $b$, it will traverse all its referenced blocks. Every time a new reference block $b'$ is discovered, it will perform a backward $BFS$ to `look back' to see if itself is already covered by the $b$'s parent block $p$. If yes, $b'$ would not be added to the forward $BFS$ queue. To avoid the complexity of the backward $BFS$, we add the previously calculated diff set to the covered set $C$, which will be passed to GetDiffSet() as a parameter. To be more specific, when a backward BFS is performed, the blocks in $C$ will not be added to the search queue. This backward search algorithm is denoted as IsCovered() and described in detail in Algorithm~\ref{algo:is_covered}.

\begin{figure}[!ht]
\begin{center}
\includegraphics[width=0.55\textwidth]{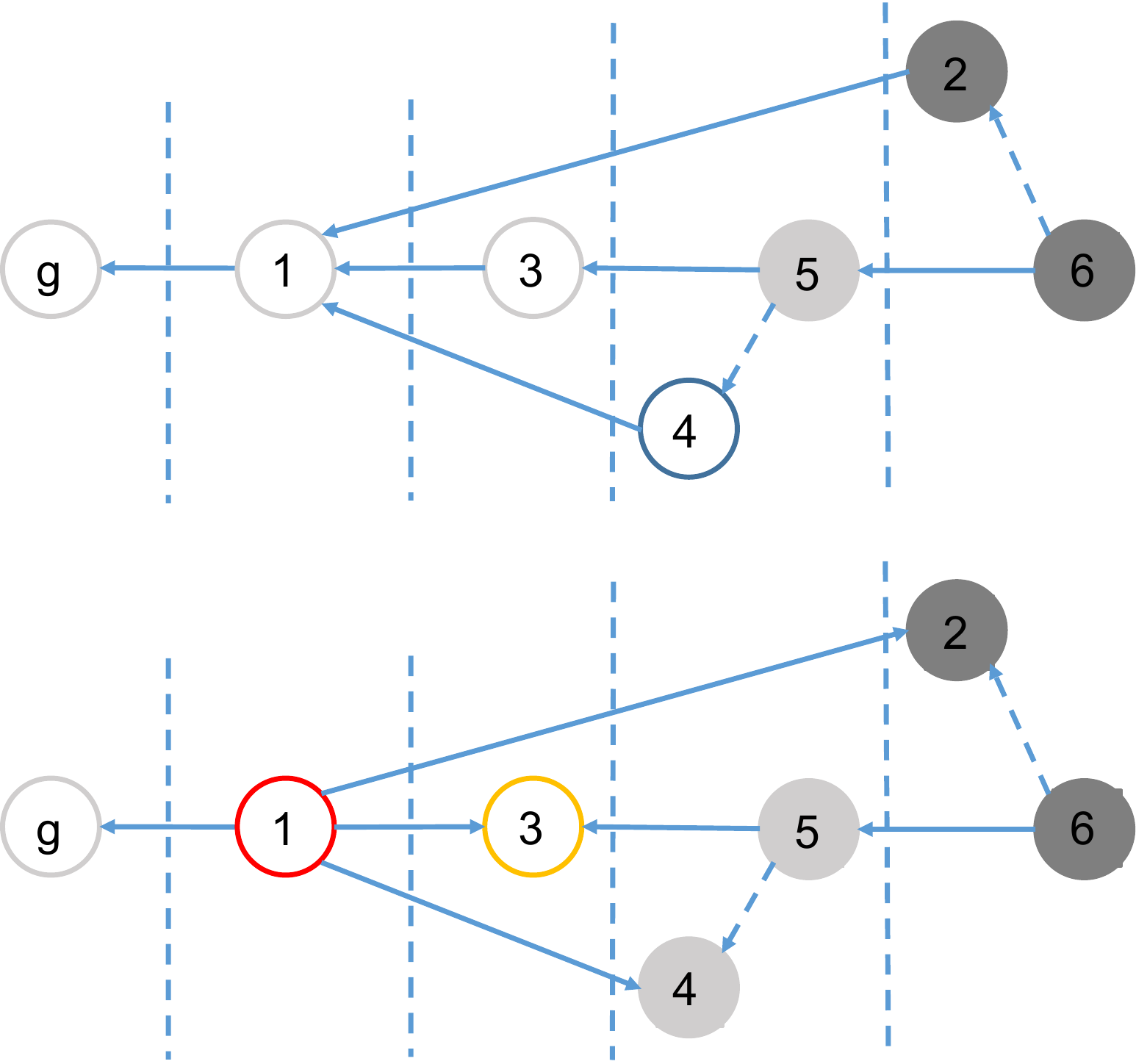}
 \caption{
  Example of the streaming get diff set method.
  }
\label{get_diff}
\end{center}
\end{figure}

Figure~\ref{get_diff} shows the example of the GetDiffSet() method for block $5$. It first performs forward BFS to find block $4$, which does not have children, then it will be added to the diff set. $4$, then move forward to $1$, which has three children. If it detects $3$, which is the parent of $5$, it will stop searching promptly. If it continues searching on $2$ or $4$, these two blocks would not be added to the search queue, because they are already in the covered set.

\subsection{Optimization of TopOrder()}
The topological order is used in sorting the blocks in the same epoch. To get the topological order, every time, there needs a top sort of the whole DAG from scratch. However, we can easily update the topological order when a new block is added or merged. The update rule is when a new block is added; its topological position will be as (1) shows. This step can be done in $O(1)$

\begin{equation}
   TopScore(G, b) \gets min(TopScore(G, Parent(b)), \\
   TopScore(G, Reference(b))) + 1
\end{equation}

To summarize, the optimized streaming operators can achieve performance improvement as Table~\ref{tab:improvement} shows.

\begin{table}[]
\caption {Analysis of Graph properties calculation} \label{tab:improvement}
\begin{center}
\begin{tabular}{|l|l|l|l|}
\hline
Graph Property  & Algorithm used & Complexity & Tot \\ \hline
Score(G, b)   & MCMC()   & $O(|B|)$    & $O(|B|)$ \\ \hline
ParentScore(G, b)  & Pivot()   & $O(|B|)$    & $O(|B|)$ \\ \hline
Past(G,b) - Past(G,p) & StreamNetOrder() & $O(|B|)$    & $O(|B|)$ \\ \hline
TopOrder(G, b)  & StreamNetOrder() & $O(|1|)$    & $O(|1|)$ \\ \hline
\end{tabular}
\end{center}
\end{table}

\subsection{Genesis Forwarding}
The above algorithm solved the problem of how to dynamically maintaining the information needed for graph computation. However, it still needs to update the information until the genesis block. With the size of the graph growing, the updating process will become harder to compute. With the growth of DAG size, the old historical confirmed blocks are being confirmed by more and more blocks, which are hard to be mutated. Furthermore, the exact probability can be computed in formula (1). Hence, we can design a strategy to forward the genesis periodically and fix the historical blocks into a total ordered chain. The criteria to forward the genesis are based on the threshold of ParentScore(). Suppose we define this threshold as $h = n-m $, then we only forward the genesis if:

\begin{equation}
   \exists b | b \in Chain(G, g), for \\
   \forall b' | b' \in \overline{Chain(G,g)}, such that \\
   ParentScore(b) > ParentScore(b') + h
\end{equation}

\begin{figure}[!ht]
\begin{center}
\includegraphics[width=0.75\textwidth]{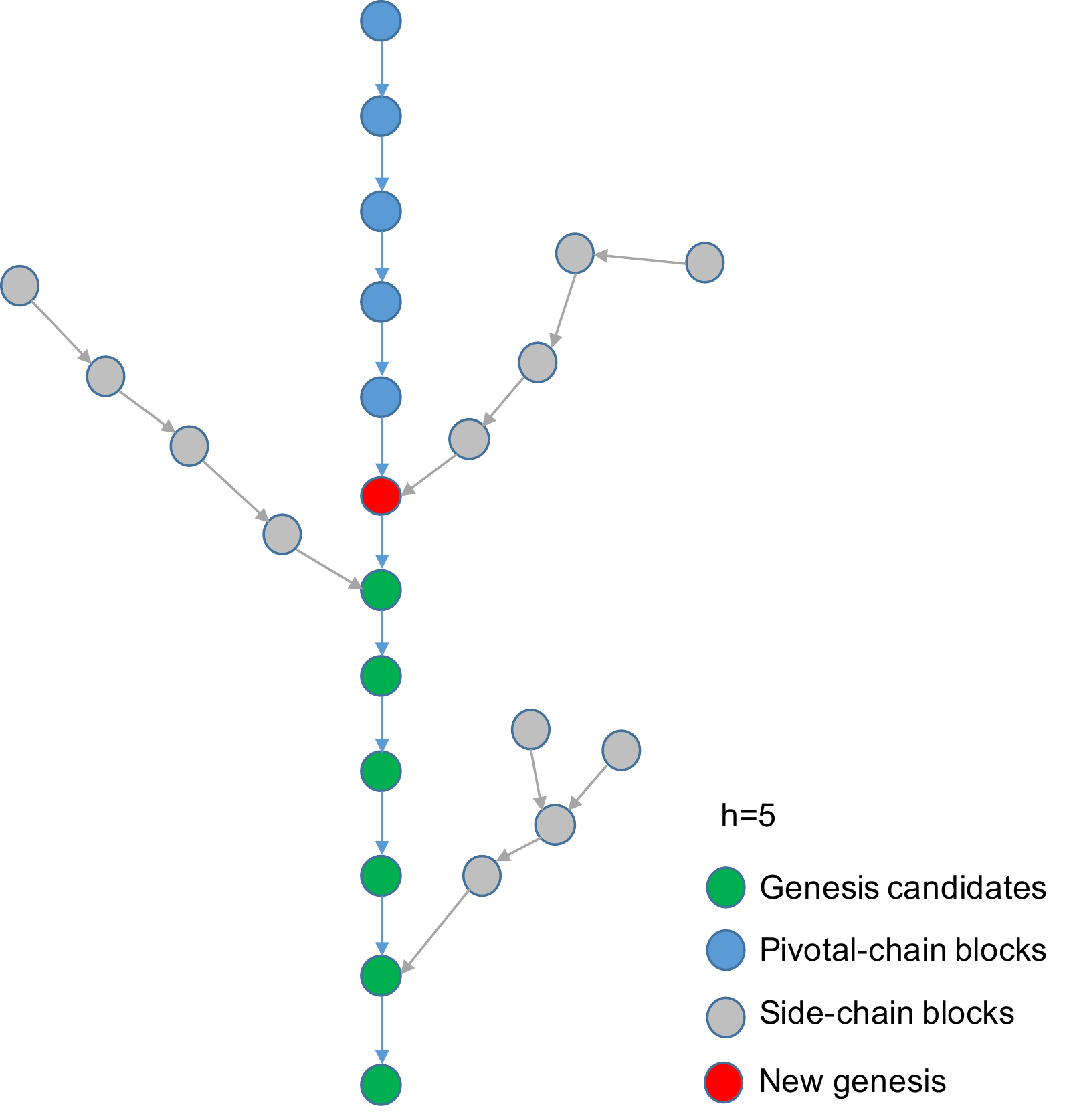}
 \caption{
  Example of genesis forward method.
  }
\label{genesis_forward}
\end{center}
\end{figure}

In Figure~\ref{genesis_forward}, we set $h=5$, and there are three side chains with $\forall b' | b' \in \overline{Chain(G,g)}$, $ParentScore(b')<=4$. And in pivot chain, there are multiple blocks $b$ that has $ParentScore(b) >= 9$, they are candidates for the new genesis, we choose the block with minimum $ParentScore$ as the new genesis.

Besides, after the new genesis has been chosen, we will induce a new DAG in memory from this genesis; furthermore, persist the `snapshot' total order (Conflux paper has the same definition, but it does not show the technical detail, we do not view it trivial) in the local database. Once the total order is queried, a total order based on the current DAG will be appended to the end of the historical snapshot total order and be returned. Also, the vertices in the UTXO graph that belongs to the fixed blocks will be eliminated from the memory and be persisted to disk as well. The algorithm is as Algorithm~\ref{algo:genesis_forward} shows.

\IncMargin{1em}
\begin{algorithm}
\SetKwData{Left}{left}\SetKwData{This}{this}\SetKwData{Up}{up}
\SetKwFunction{Union}{Union}\SetKwFunction{FindCompress}{FindCompress}
\SetKwInOut{Input}{input}\SetKwInOut{Output}{output}

\KwIn{ Graph $G=<B, g, P, E>$ }

    \While {Node is running}{
        \uIf{$\exists b$ satisties (3)}{
            $ O = TopOrder(G, g) $\;
            $g' \gets b$ \;
            $G' \gets induceGraph(G, g') $ \;
            $ pS = ParentScore(G', g') $\;
            $ S = Score(G', g') $\;
            $ O' = TopOrder(G', g') $\;
            $G \gets G'$ \;
            persist $O - O'$ \;
        } 
        sleep ($t$) \;
    }

\caption{{ Genesis Forward Algorithm.}}
\label{algo:genesis_forward}
\end{algorithm}
\DecMargin{1em}

\subsection{The Direct Signal Gossip Protocol}
There are solutions in \cite{demers1988epidemic} to minimize the message passing in the gossip network. Moreover, in Hyperledger \cite{androulaki2018hyperledger} they have adopted the PUSH and PULL model for the gossip message propagation. However, their system is aiming at permissioned chain. Suppose the size of the hash of a block is $H$, we designed the direct signal algorithm. The algorithm is divided into two steps, once a node generates or receives a block, it firstly broadcast the hash of the block, this is the PUSH step. Once a node receives a hash or a set of a hash, it will pick one source of the hash for the block content, and this is the PULL step. The direct signal algorithm's complexity will be $O(LH + NB)$ and for a node averaged to $O(\frac{LH}{N} + 1)$ The algorithm is as Algorithm~\ref{algo:gossip} shows.

\IncMargin{1em}
\begin{algorithm}
\SetKwData{Left}{left}\SetKwData{This}{this}\SetKwData{Up}{up}
\SetKwFunction{Union}{Union}\SetKwFunction{FindCompress}{FindCompress}
\SetKwInOut{Input}{input}\SetKwInOut{Output}{output}

\KwIn{ Graph $G=<B, g, P, E>$ }

    \While {Node is running}{
        \uIf{Generate block $b$}{
            Broadcast $b$ to neighbors \;
        } 
        \uIf{Receive block $b$}{
            $h \gets Hash(b)$ \;
            $cache[h] \gets b$ \;
            Broadcast $h$ to neighbors \;
        } 
        \uIf{Received request $h$ from neighbor $n$ }{
            $b \gets cache[h]$ \;
            Send $b$ to $n$ \;
        } 
        \uIf{Received hash $h$ from neighbor $n$}{
            $b \gets cache[h]$ \;
            \uIf{ $b = NULL $ } {  
                Send request $h$ to $n$ \;
            }
        } 
    }

\caption{{ The Direct Signal Gossip Algorithm.}}
\label{algo:gossip}
\end{algorithm}
\DecMargin{1em}


\section{Experimental Results}
  
\subsection{Implementation}

\begin{figure}[!ht]
\begin{center}
\includegraphics[width=0.75\textwidth]{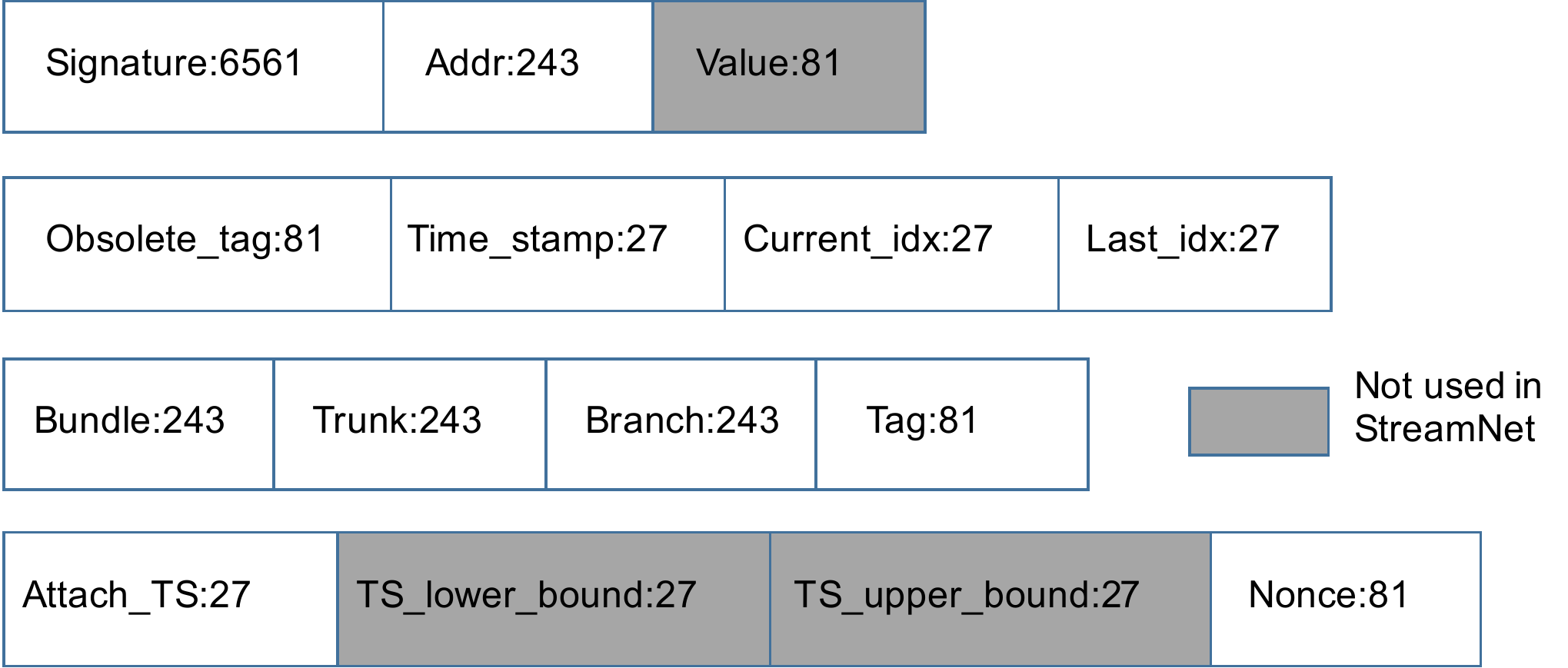}
  \caption{
    Block header format, the main transaction information is stored in the signature part. The addr is sender's address, the timestamp is the time the block has been created, current/last index and the bundle is used for storing the bundle information, trunk and branch are the hash address to store the parent and reference location, the tag is used for store some tagging information, addtach\_TS is when the block is attached to the StreamNet, the nonce is used in POW calculation.
   }
\label{block_header}
\end{center}
\end{figure}

We have implemented the StreamNet based on the IOTA JAVA reference code (IRI) v1.5.5 \cite{IOTACode}.
We forked the code and made our implementation; the code is freely available at \cite{StreamNet}. In this paper, we use version v0.1.4-streamnet in the v0.1-streamnet beta branch.

\begin{itemize}
  \item The features we have adopted from the IRI are:
  \begin{itemize}
    \item The block header format, as shown in Figure~\ref{block_header}. Some of the data segments are not used in StreamNet, which are marked grey.
    \item Gossip network, the network is a bi-directional network in which every node will send and receive data from its peers;
    \item Transaction bundle, because of the existence of the bundle hash feature, StreamNet can support both the single transaction for a block and batched transactions as a bundle.
    \item Sponge hash functions, which is claimed to be quantum immune, in our experiment, the POW hardness is set to 8, which is the same as the testnet for IOTA.
  \end{itemize}

  \item The features we have abandoned from the IRI are:
  \begin{itemize}
    \item The iota's transaction logic including the ledger validation part;
    \item The milestone issued by coordinators, which is a centralized setup.
  \end{itemize}

  \item The features we have modified based on the IRI is:
  \begin{itemize}
    \item The tip selection method based on MCMC, since the tip selection on IRI has to find a milestone to start searching, we replace this with a block in the pivotal chain instead.
  \end{itemize}

  \item The features we have added into the StreamNet are:
  \begin{itemize}
    \item The consensus algorithms, and we have applied the streaming method directly in the algorithms;
    \item The UTXO logic stored in the signature part of the block header used the graph data structure to store UTXO as well.
    \item In IOTA's implementation, the blocks are stored in the RocksDB \cite{RocksDB} as the persistence layer, which makes it inefficient to infer the relationships between blocks and calculate graph features. In our implementation, we introduced an in-memory layer to store the relationships between blocks, such that the tip selection and total ordering algorithm will be accelerated.
  \end{itemize}
\end{itemize}

\subsection {Environment Set Up}

\begin{figure*}[!ht]
\begin{center}
\includegraphics[width=0.95\textwidth]{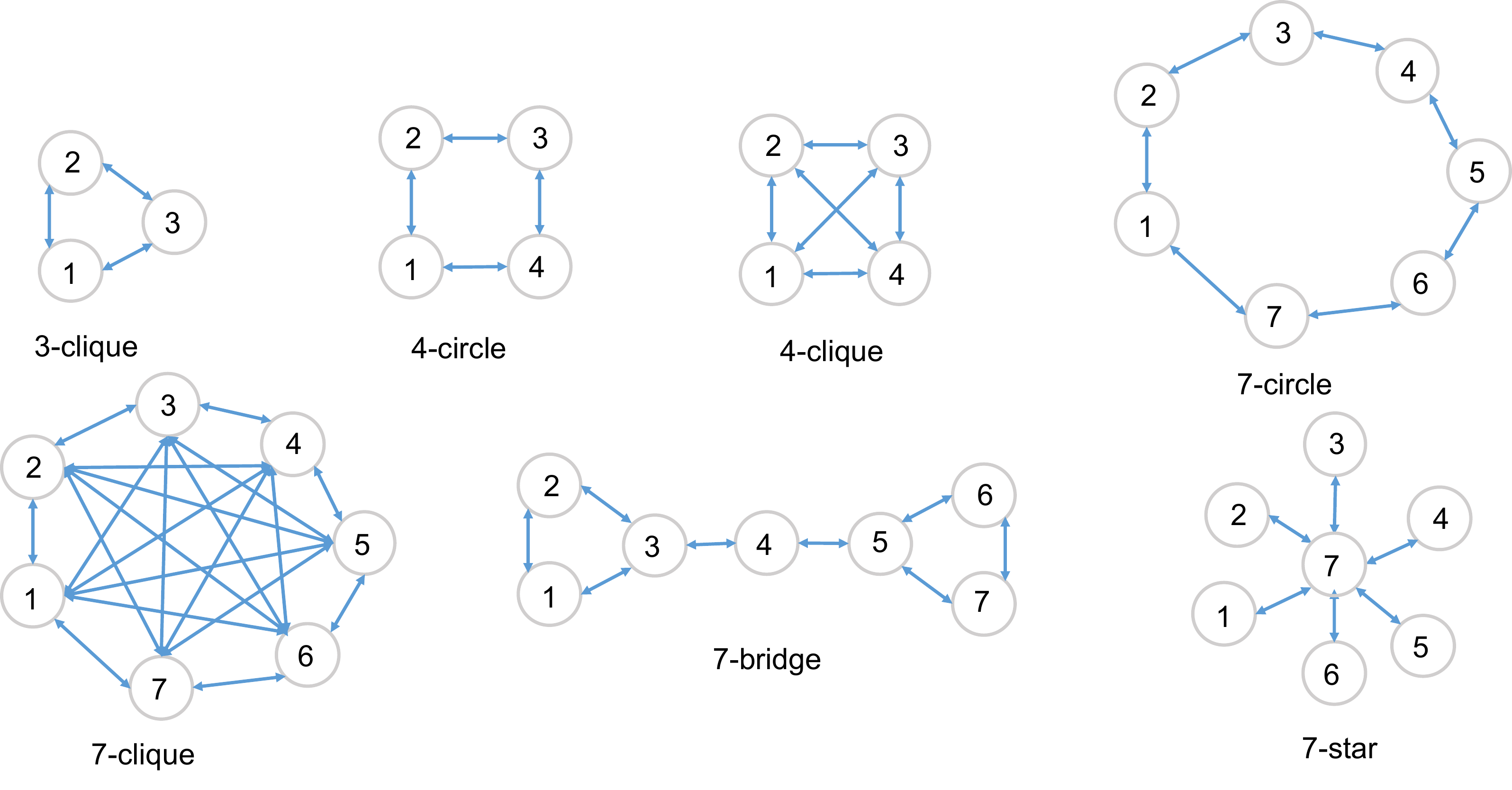}
  \caption{
    Cluster set up for different network topologies.
   }
\label{cluster_set_up}
\end{center}
\end{figure*}

We have used the AWS cloud services with 7 virtual machines, for each node, it includes a four-core AMD EPYC 7571, with 16 Gb of memory size and 296Gb of disk size.
The JAVA version is 1.8, we have deployed our service using docker, and the docker version is 18.02.0-ce.

We have 7 topologies set up of nodes, which are shown in Figure~\ref{cluster_set_up}, these configurations are aiming to test:
\begin{itemize}
  \item The performance when the cluster connectivity is high (congestion of communications, like 3-clique, 4-clique, 7-clique, and 7-star);
  \item The performance when the cluster diameter is high (long hops to pass the message, like 4-circle, 7-circle, 7-bridge);
\end{itemize}

As for the data, we have created 1,000 accounts, with the genesis account having 1,000,000,000 tokens in the coinbase block. We divided the accounts into two groups (each group will have 500 accounts), the first group will participate in the ramp-up step, which means the genesis account will distribute the tokens to these accounts. Moreover, for comparison, we have issued four sets of different size transactions (5000, 10000, 15000, and 20000), respectively. In the execution step, the first group of accounts will issue transactions to the second group of accounts, which constructs a bipartite spending graph. Since there are more transactions than the number of accounts, there will be double-spend manners in this step. The number of threads in this procedure is equal to the number of nodes for each configuration. Jmeter \cite{halili2008apache} is utilized as the driver to issue the transactions, and Nginx \cite{nedelcu2010nginx} is used to evenly and randomly distribute the requests to different nodes.

\subsection {Results and Discussions}

\subsubsection {Block generation rate test}
\begin{figure}[!ht]
\begin{center}
\includegraphics[height=0.75\textwidth, width=0.75\textwidth]{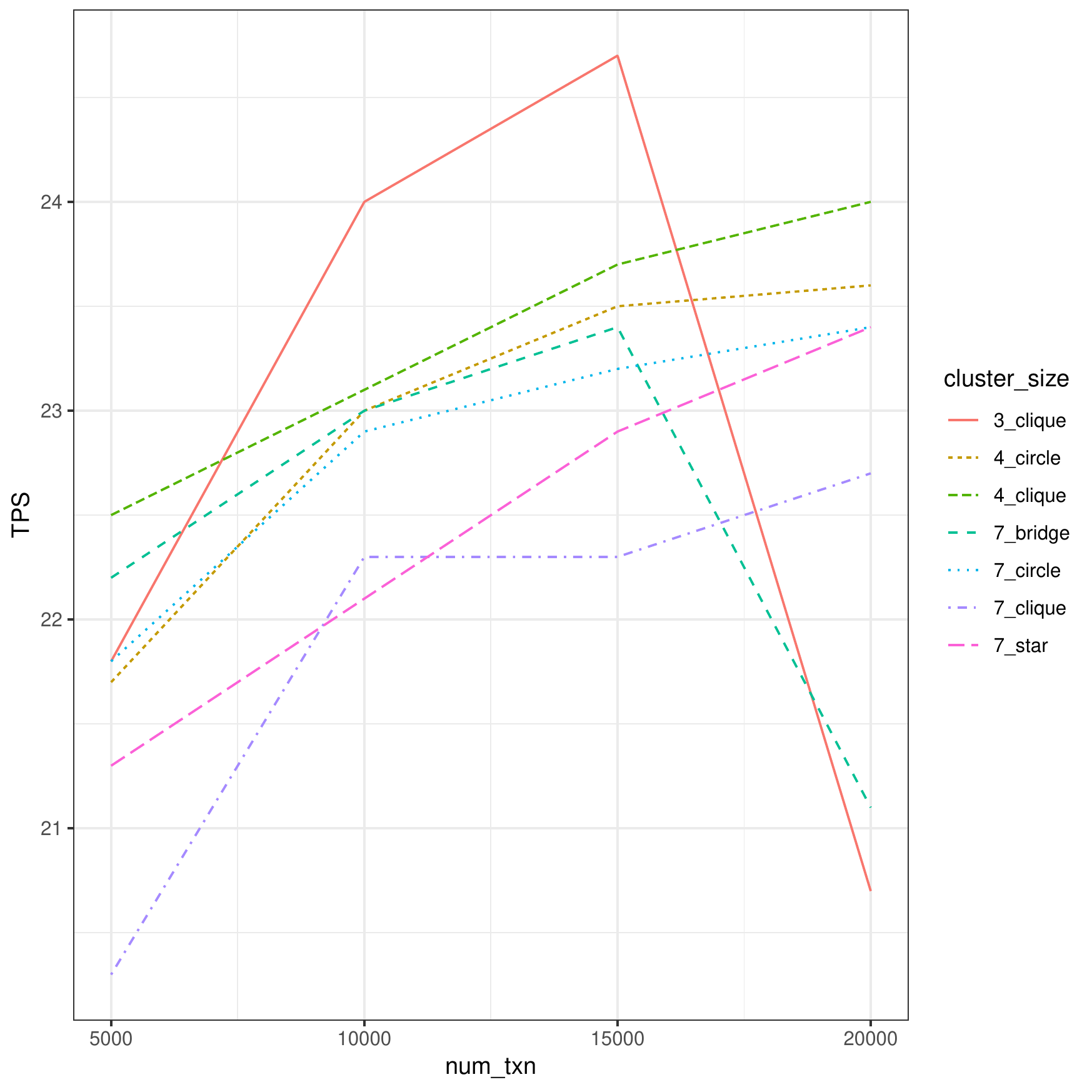}
  \caption{
    Experimental results for block generation rate.
   }
\label{single_txn}
\end{center}
\end{figure}

To test the block generation rate, we set each block in StreamNet to have only one transaction. Furthermore, the performance on this configuration is as Figure~\ref{single_txn} shows. First, as the size of the cluster grows, the network will witness little performance loss on all of the data scales. In the experiment, we can also see that with the growth of the data, the average TPS on most of the configurations have grown a little bit (some outliers need our time to triage),
this is because the genesis forwarding algorithm needs some ramp-up time to get to the stable growth stage. Considering the system is dealing with a growing graph instead of a chain and the complexity analysis in the previous section, the experiment clearly shows that our streaming algorithm sheds light on how to deal with the growing DAG.

\subsubsection {Bundle transaction test}

\begin{figure}[!ht]
\begin{center}
\includegraphics[height=0.75\textwidth, width=0.75\textwidth]{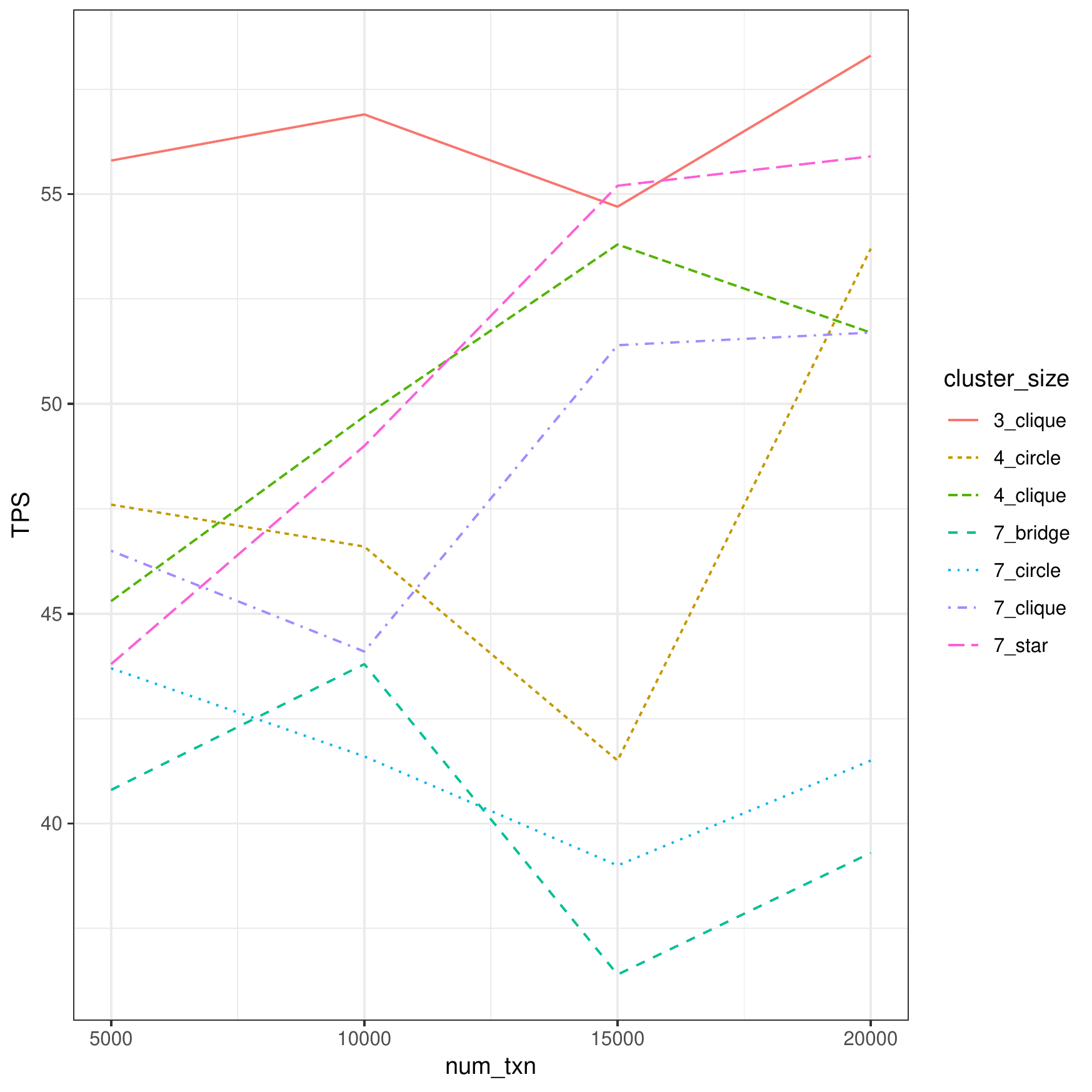}
  \caption{
    Experimental results for bundle transaction.
   }
\label{multi_txn}
\end{center}
\end{figure}

By default, each block in StreamNet will support bundle transactions.
We set each bundle to contain 20 transactions, and for each block, there are approximately 3 transactions included. The performance on this configuration is as Figure~\ref{multi_txn} shows. In this experiment, we can see that the performance (TPS) comparing with the block test improved more than twice. This is because there will be less POW works to be done. Besides, with the growth of the data, we do not witness a noticeable performance downturn. Nevertheless, there are some performance thrashing in the experiment, which needs more study.

\section{Conclusion}
In this paper, we proposed a way to compute how to grow the blocks in the growing DAG based blockchain systems. 
And how to maintain the total order as the DAG structure is dynamically turning larger.
We referred one of the earliest DAG implementation IRI to conduct our own experiments on clusters of different size and topology. 
Despite the network inefficiency in the IRI implementation, 
our method is proven to be able to tolerate the increasing complexity of the graph computation problems involved. 
This is due to the streaming graph computing techniques we have introduced in this paper.

%
%

\bibliographystyle{unsrt}
\bibliography{template}

\end{document}